\input{aipcheck}

\documentclass[
  ,draft            
  ]
  {aipproc}
\usepackage{amsmath}

\layoutstyle{8x11single}

\begin{document}

\newcommand{\PA}{Pad\'{e} approximant}
\newcommand{\GE}{gradient expansion}
\newcommand{\KS}{Kohn--Sham}
\newcommand{\TF}{Thomas--Fermi}
\newcommand{\HF}{Hartree--Fock}
\newcommand{\HLM}{Hooke's law model}
\newcommand{\HLMTEA}{Hooke's law model for two-electron atoms}
\newcommand{\DFT}{density functional theory}

\newcommand{\eref}[1]{equation~(\ref{#1})}
\newcommand{\eeref}[2]{equations~(\ref{#1}) and (\ref{#2})}

\title{Convergent sum of gradient expansion of the kinetic-energy density functional up to the sixth order term using \PA}


\author{Alexey Sergeev }{
  address={Qatar Environment and Energy Research Institute,
  PO Box 5825, Doha, Qatar}
}

\author{Raka Jovanovic}{
  address={Qatar Environment and Energy Research Institute,
  PO Box 5825, Doha, Qatar}
}

\author{Sabre Kais}{
  address={Qatar Environment and Energy Research Institute,
  PO Box 5825, Doha, Qatar}
}

\author{ahhad H Alharbi}{
  address={Qatar Environment and Energy Research Institute,
  PO Box 5825, Doha, Qatar}
}

\begin{abstract}
The \GE\ of the kinetic energy functional, when applied for atoms or finite systems, usually grossly overestimates the energy in the fourth order and generally diverges in the sixth order. We avoid the divergence of the integral by replacing the asymptotic series including the sixth order term in the integrand by a rational function. \PA s show moderate improvements in accuracy in comparison with partial sums of the series. The results are discussed for atoms and \HLMTEA.
\end{abstract}

\maketitle
\section{Introduction}


For all electronic systems, Hohenberg and Kohn \cite{H01} proved the existence of a ``universal  functional'' that accounts for the total energy except those contributions due to an external potential (mainly due to the electron-nuclei Coulombic interaction). The main challenge is that the universal functional is not known and thus Kohn and Sham \cite{K01} devised an alternative approach to approximate the ``universal  functional''. In this approach, the kinetic term for non-interacting electrons was treated exactly but the exchange-correlation term are left unknown. Furthermore, in their initial work, Hohenberg and Kohn \cite{H01} suggested that the components of the ``universal  functional''; namely, the kinetic, exchange, and correlation contributions, can be expanded using gradient based expansions. Since then, many approximations have been developed. For kinetic energy, the starting point has been mostly \TF\ functional. In the limit of infinitely extended uniform density, the kinetic energy $T_\mathrm{s}[\rho]$ is simply \TF\ kinetic energy $T_\mathrm{TF}[\rho]$ (in density functional form) \cite{L02}. For slowly varying densities, corrections to $T_\mathrm{TF}$ can be derived using gradient expansions. Different approaches have been used and they agrees in most cases. The most frequently used procedures are Kirzhnits expansion \cite{K02,H02,M01}, partition-function approach \cite{H02}, and Grammaticos-Voros algebraic method \cite{G01}. As a result, the kinetic energy density functional (KEDF) can be arranged in the following series:
\begin{equation}
T_\mathrm{s}[\rho(\textbf{r})]= T_0+ T_2+ T_4+ T_6+ \ldots,
\label{gexp}
\end{equation}
where $T_0= T_\mathrm{TF}$, which will be presented in the next section, $T_n= \int \tau_n d\textbf{r}$, and the kinetic energy densities $\tau_n$ collect terms that have gradients of the total order $n$. With the exception of solid state systems \cite{Y01}, the sixth-order term of the \GE\ generally diverges for a finite system. To eliminate the divergence, many additional approximations have been suggested. Pearson and Gordon \cite{P01} suggested local truncation of the gradient expansion with additional corrections to avoid vanishing densities. Also, many alternative approximate re-summation methods \cite{L01} including \PA s \cite{V01,C01} were suggested. The suggested \PA s are mostly based on $\tau_2/\tau_0$ expansions to resemble different asymptotic behaviors \cite{V01,C01}. Obviously, each approach has advantages and disadvantages and each has a range of applicability.

In this paper, we present a systematic approach that allows to avoid regions where higher order terms are relatively large and are no longer ``small corrections'' while taking advantage of these terms in areas where they are small and improve the accuracy. So, we use a \PA\ to have a convergent sum of gradient expansion approximation up to the sixth-order term. In many cases, adding of the correction $T_4[\rho]$ usually makes the result less accurate, because this correction is valid only asymptotically for heavy atoms. Moreover, since $T_6[\rho]$ diverges for finite systems, the knowledge of $\tau_6$ is assumingly -from the first glance- useless. By using the proper \PA , it is possible to ``sum'' approximately the \GE\ of kinetic energy so that the approximant has some asymptotic behavior, but suppresses divergent terms in a region away from asymptotics.

\section{The method}

As aforementioned, in the limit of uniform density \cite{L02,L03}, the kinetic energy $T_\mathrm{s}[\rho]$ is
\begin{equation}
T_\mathrm{TF}= C_\mathrm{TF} \int\rho^{5/3}\, d\textbf{r},\; C_\mathrm{TF}= \frac{3}{10} \left(3 \pi^2\right)^{2/3}
\label{etf}
\end{equation}
Any effect due to slowly varying densities can be accounted for approximately as corrections to Eq. \ref{etf} using gradient expansions. So, the kinetic energy density functional can be organized in a series as in Eq. \ref{gexp}. Using Kirzhnits expansion approach, Kirzhnits \cite{K02}, Hodges \cite{H02}, and Murphy \cite{M01} have derived the second-, forth-, and sixth-order terms respectively. These terms are
\begin{equation}
\tau_2 =\frac1{72}\frac{\left(\nabla \rho(\vec{r})\right)^2}{\rho(\vec{r})}, \,\,\,\,\,\,\,\,\,\,\,\,\,\,\,\,\,\
\tau_4 = \frac{\left(3 \pi^2\right)^{-2/3}}{540}\rho^{1/3} \left[\left(\frac{\nabla^2\rho}{\rho}\right)^2 -\frac98 \left(\frac{\nabla^2\rho}{\rho}\right)\left(\frac{\nabla\rho}{\rho}\right)^2+ \frac13 \left(\frac{\nabla\rho}{\rho}\right)^4\right].
\label{tau4}
\end{equation}
and
\begin{multline}
\tau_6 = \frac{\left(3 \pi^2\right)^{-4/3}}{45\,360}\rho^{-1/3} \left[
13 \left(\frac{\nabla\nabla^2\rho}{\rho}\right)^2+ \frac{2575}{144}\left(\frac{\nabla^2\rho}{\rho}\right)^3+
\frac{249}{16}\left(\frac{\nabla\rho}{\rho}\right)^2\left(\frac{\nabla^4\rho}{\rho}\right)\right.
 +\frac{1499}{18}\left(\frac{\nabla\rho}{\rho}\right)^2\left(\frac{\nabla^2\rho}{\rho}\right)^2
\\-\frac{1307}{36}\left(\frac{\nabla\rho}{\rho}\right)^2\left(\frac{\nabla\rho\cdot \nabla\nabla^2\rho}{\rho^2}\right)
+\frac{343}{18}\left(\frac{\nabla\rho\cdot \nabla\nabla\rho}{\rho^2}\right)^2
 +\left.\frac{8341}{72}\left(\frac{\nabla^2\rho}{\rho}\right)\left(\frac{\nabla\rho}{\rho}\right)^4
-\frac{1\,600\,495}{2592}\left(\frac{\nabla\rho}{\rho}\right)^6\right].
\label{tau6}
\end{multline}
Generally, the sixth-order term diverges for atoms and molecules. Actually, it diverges when the density vanishes at any position. A study of the \GE\ for atomic isoelectronic series \cite{W01} shows that that $T_0+T_2$ is more accurate approximation than $T_0+T_2+T_4$ in almost all cases, except for highly ionized Ar-like ions. A similar study for all neutral atoms \cite{M02} shows that $T_0+T_2$ is more accurate than $T_0+T_2+T_4$ for all atoms up to potassium. Only for heavier atoms, the situation reverses. It rises the question of whether the forth order term $T_4$ and the diverging sixth-order density $\tau_6$ could be of any use for the case of finite systems. A typical behavior of the gradient correction for an atomic system that in a narrow region of $r$ we have the \GE\ seems convergent, i.e. $|\tau_6|< |\tau_4|< |\tau_2|< |\tau_0|$. However, everywhere else $\tau_6$ is no longer a small correction to the lower order terms. For sufficiently large $r$, we have an opposite inequality $|\tau_6|> |\tau_4|> |\tau_2|> |\tau_0|$, that means that all corrections become erroneous. So, the objective is to ``sum'' the \GE\ at each point $r$ by replacing it by another function that has some asymptotic behavior, but suppresses the divergent terms in a region away from asymptotics. Let's consider a family of densities parameterized by a dummy variable $g>0$,
\begin{equation}
\rho_g(\vec{r})= g \rho(r),
\label{rhog}
\end{equation}
so that the original density $\rho$ can be recovered from this family of functions by setting $g=1$.
Then, the \GE\ for the density can be organized as in the following series
\begin{equation}
\tau_g(\vec{r})= g^{5/3} \left( \tau_0(\vec{r}) + g^{-2/3}\tau_2(\vec{r}) + g^{-4/3}\tau_4(\vec{r}) + g^{-6/3}\tau_6(\vec{r}) + \ldots \right).
\label{taug}
\end{equation}
The expression in parenthesis in Eq. \ref{taug} is obviously a power series in $x= g^{-2/3}$. So, asymptotically for small $x$ we have
\begin{equation}
f(x)= \tau_0 + \tau_2 x+ \tau_4 x^2+ \tau_6 x^3+ \ldots,
\label{xexpans}
\,\,\,\,\,\,\,\,\,\,\ where  \,\,\,\,\,\,\,\,\,\,\
f(x)= g^{-5/3} \tau_g(\vec{r}).
\end{equation}
We would like to replace $f(x)$ by an approximation $\tilde{f}(x)$ having the same asymptotic expansion as $f(x)$ given by Eq. \ref{xexpans}. We seek to have $\tilde{f}(1)\approx \tau_0 + \tau_2 + \tau_4 + \tau_6 $ if $|\tau_6|< |\tau_4|< |\tau_2|< |\tau_0|$ and $\tilde{f}(1)\approx \tau_0 + \tau_2 $ if $|\tau_6|$ is large. The \PA\ $[2/1]$ to the series Eq. (\ref{xexpans}) is defined as a ratio of two polynomials of degree 2 and 1 in $x$,
\begin{equation}
f^{[2/1]}(x)= \frac{a_0+a_1 x+a_2 x^2}{b_0+b_1 x}
\label{f21def}
\end{equation}
with $b_0\ne 0$ that has the same expansion in $x$ up to the third order[$f^{[2/1]}(x)= f(x)+ o(x^3)$]. Explicitly
\begin{equation}
f^{[2/1]}(x)= \tau_0+ \tau_2 x+ \frac{\tau_4^2 x^2}{\tau_4- \tau_6 x}.
\label{f21}
\end{equation}
Notice that if $\tau_6\rightarrow \infty$, then $f^{[2/1]}(1)= \tau_0+ \tau_2$. We define $\tau^{[2/1]}= f^{[2/1]}(1) = \tau_0 + \tau_2+ \frac{\tau_4^2}{\tau_4-\tau_6}$ and $T^{[2/1]}= 4\pi \int_0^{\infty}r^2\tau^{[2/1]}\,dr$ where the later equation is written in case of spherical symmetry. If the integrand has poles, i.e. $\tau_4=\tau_6$ for some $r$, then the integral can be defined through Cauchy principal value. In similar way, we consider here the \PA\ $\tau^{[1/1]}= \tau_0 + \frac{\tau_2^2}{\tau_2-\tau_4}$ along with the partial sum $\tau_0+\tau_2+\tau_4$. Even if the integral of this partial sum converges, the convergence is slow because of behavior $\sim \rho^{1/3}$ for large $r$. For the \PA\ $\tau^{[1/1]}$, much faster convergence $\sim\rho^{7/3}$ is expected as obtained and presented in the next section.

\section{Results and discussions}

The Hamiltonian for the \HLMTEA\ is given by
\begin{equation}
H =-\frac12(\nabla_1^2+ \nabla_2^2)+ \frac{\omega^2}{2} (r_1^2+ r_2^2)+ \frac1{r_{12}}.
\label{hkh}
\end{equation}
The ground state wavefunction is known analytically for some values of $\omega$. For $\omega=1/2$, it is \cite{K03,K04}
\begin{equation}
\Psi(r_1, r_2) =N_0 \left(1+\dfrac12 r_{12}\right)\, e^{-(1/4)(r_1^2+r_2^2)},
\,\,\,\,\,\, where \,\,\,\,\,\,\,\, 
N_0 =\left[4\pi^{5/2} \left(8+5\pi^{1/2}\right)\right]^{-1/2}.
\label{hkn}
\end{equation}
and the density is (corrected Eq. (7) from \cite{K03})
\begin{equation}
\rho(r) =N_0^2 \,e^{-\frac12 r^2} \left\{\left(\frac{\pi}2\right)^{1/2} \left[\frac74+ \frac14 r^2+\left(r+\frac1{r}\right) \mathrm{erf}(2^{-1/2} r)\right]
+e^{-\frac12 r^2}\right\}.
\label{hkrho}
\end{equation}
If $\omega\ne 1/2$, then we used a variational approach with the trial wavefunction in the form similar to Hylleraas variational function for helium \cite{S02,S03}
\begin{equation}
\Psi(\textbf{r}_1,\textbf{r}_2)= \sum_{i,j,k\ge 0,\,i+j+k=N}
C_{ijk} \left(r_1^i r_2^j e^{-\alpha_1 r_1-\alpha_2 r_2}+
r_2^i r_1^j e^{-\alpha_1 r_2-\alpha_2 r_1}\right) r_{12}^k
\label{psitrial}
\end{equation}
with $\alpha_1=2\sqrt{\omega}$, $\alpha_2=3.2\sqrt{\omega}$ (ideally, the values of $\alpha_1$ and $\alpha_2$ should be chosen by minimizing the energy, but since it involves non-linear equations, we did it numerically).  The results are shown in Table \ref{tabl1} For $\omega< 1$, it appears that the \PA\ [2/1] is the most accurate, and for $\omega\ge 1$ the partial sum $T_0+T_2$ is the best.
 \setlength{\tabcolsep}{10pt}
\renewcommand{\arraystretch}{1.2}
\begin{table}[hpt]
\begin{tabular}{c c c c c c c}
\hline
\multicolumn{2}{c}{ }& \multicolumn{5}{c}{Percent error of} \\
\multicolumn{1}{c}{$\omega$}& \multicolumn{1}{c}{$T_\mathrm{s}$}& \multicolumn{1}{c}{$T_0$}& \multicolumn{1}{c}{$T_0+T_2$}& \multicolumn{1}{c}{$T_0+T_2+T_4$}& \multicolumn{1}{c}{$T^{[1/1]}$}& \multicolumn{1}{c}{$T^{[2/1]}$}\\[2pt]
\hline
1/4& 0.30036& -12.7& -1.67& 15.6& 0.48& -1.15\\
1/2& 0.63525& -11.9& -0.78& 16.5& 1.27& -0.26\\
1& 1.32757& -11.3& -0.19& 15.4& 1.81& 0.33\\
4& 5.62884& -10.7& 0.45& 15.1& 2.4& 0.98\\
\hline
\end{tabular}
\caption{Comparison of kinetic energy obtained from density $\rho$ by summation of gradient expansion using different methods, for \HLM.}
\label{tabl1}
\end{table}
\renewcommand{\arraystretch}{1.0}
\setlength{\tabcolsep}{10pt}
\renewcommand{\arraystretch}{1.2}
\begin{table}
\label{tabl2}
\begin{tabular}{c c c c c c c}
\hline
\multicolumn{2}{c}{ }& \multicolumn{5}{c}{Percent error of} \\
\multicolumn{1}{c}{Atom}& \multicolumn{1}{c}{$T_\mathrm{HF}$ (a.u.)}& \multicolumn{1}{c}{$T_0$}& \multicolumn{1}{c}{$T_0+T_2$}& \multicolumn{1}{c}{$T_0+T_2+T_4$}& \multicolumn{1}{c}{$T^{[1/1]}$}& \multicolumn{1}{c}{$T^{[2/1]}$}\\[2pt]
\hline
He& 2.8617& -10.5& 0.59& 3.57& 2.01& 0.53\\
Li& 7.4328& -10.1& 0.62& 3.09& 2.00& 0.61\\
Be& 14.573& -9.9& 0.50& 2.9& 1.86& 0.53\\
Ne& 128.55& -8.4& -0.55& 0.95& 0.50& -0.51\\
Ar& 526.81& -7.0& -0.49& 0.69& 0.32& -0.43\\
Kr& 2752.05& -5.9& -0.69& 0.18& -0.09& -0.63\\
Xe& 7232.13& -5.2& -0.68& 0.07& -0.17& -0.62\\
\hline
\end{tabular}
\caption{Comparison of kinetic energy obtained from density $\rho$ by summation of gradient expansion using different methods, for atoms.}
\end{table}
\setlength{\tabcolsep}{6pt}
\renewcommand{\arraystretch}{1.0}
For the atoms, we calculated the densities using \HF\ wavefunctions from Clementi Atomic Data Tables \cite{C02}. The results of summation of the \GE\ are given in Table \ref{tabl2}. For He, we found that the gradient expansion and its summations are similar to the case of \HLMTEA, but the sixth-order term diverges even faster. As a result, the \PA\ improves the partial sum $\tau_0+\tau_2$ only in a narrow range of $r\sim 0.6$, where $\tau_6$ is comparable with $\tau_0$. Outside of this range, $\tau^{[2/1]}\approx \tau^{(2)}$. Therefore, results of integration of these functions are very close, but still the \PA\ gives some small improvement in comparison with the partial sum. While for Li and Be, we found that the gradient expansion and its summation is similar to the case of helium atom, but the sixth-order term diverges much faster. As a result, the \PA\ does not improve accuracy in comparison with the partial sum. For heavier noble-gas atoms (Ne, Ar, Kr, and Xe), we found that the \PA\ [2/1] is still slightly more accurate than the sum of two terms, but the \PA\ [1/1] is even more accurate. Calculations show that in these cases $\tau_4< \tau_2$ in the most important area of integration when $\tau(r)> 0.01$. It explains the good accuracy of the \PA\ [1/1] which is based on fourth order expansion. However, we found that $\tau_6(r) > \tau_4(r)$ for almost all $r$, which is probably the reason of relatively poor performance of the \PA\ [2/1] that is based on sixth order expansion.

\section{Conclusion}
Here, we take advantage of the known higher order corrections to the density of the kinetic energy, $\tau_4$ and $\tau_6$, to improve the accuracy of the \GE. Simple adding of the correction $T_4= \int \tau_4(\textbf{r})\, d\textbf{r}$ usually makes the result less accurate, because this correction is valid only asymptotically for heavy atoms. Moreover, since $T_6= \int \tau_6(\textbf{r})\, d\textbf{r}$ is given by a divergent integral for finite systems, the knowledge of $\tau_6$ is assumingly useless, from the first glance. In this paper, we present a systematic approach that allows to avoid regions where higher order terms are relatively large and are no longer ``small corrections'' while taking advantage of these terms in areas where they are small and improve the accuracy. We expect that this approximation can improve the accuracy in case when there exists a significant region of $\vec{r}$ where $|\tau_6(\vec{r})|< |\tau_4(\vec{r})|$. We found that for atoms, this region is quite narrow, presumably because of the presence of Coulomb singularity. In this case the method has only small advantage over the method of plane summation. Our second example is \HLMTEA, where the Coulomb attraction $-Z/r$ is replaced by a harmonic binding $\frac{\omega^2}{2}r^2$. In the latter case, we found that the method works only in case of sufficiently weak binding, $\omega\le 1/2$. So, we expect this approximation to work very well for confined harmonic potentials such the case of treating quantum dots.

\bibliographystyle{aipproc} 
\bibliography{PadeAppr}



\end{document}